\documentclass[pra,aps,showpacs,twocolumn,superscriptaddress]{revtex4}

\usepackage{amsmath,amsfonts,amssymb,color,epsfig,graphics,graphicx,latexsym,revsymb,theorem,verbatim}

\usepackage[german]{babel}

\newtheorem{definition}{Definition}
\newtheorem{proposition}[definition]{Proposition}
\newtheorem{lemma}[definition]{Lemma}

\newtheorem{theorem}[definition]{Theorem}
\newtheorem{corollary}[definition]{Corollary}
\newtheorem{conjecture}[definition]{Conjecture}

\newtheorem{remark}[definition]{Remark}
\newtheorem{example}[definition]{Example}

\def\squareforqed{\hbox{\rlap{$\sqcap$}$\sqcup$}}
\def\qed{\ifmmode\squareforqed\else{\unskip\nobreak\hfil
\penalty50\hskip1em\null\nobreak\hfil\squareforqed
\parfillskip=0pt\finalhyphendemerits=0\endgraf}\fi}
\def\endenv{\ifmmode\;\else{\unskip\nobreak\hfil
\penalty50\hskip1em\null\nobreak\hfil\;
\parfillskip=0pt\finalhyphendemerits=0\endgraf}\fi}
\newenvironment{proof}{\noindent \textbf{{Proof.~} }}{\qed}

\def\bcj{\begin{conjecture}}
\def\ecj{\end{conjecture}}
\def\bcr{\begin{corollary}}
\def\ecr{\end{corollary}}
\def\bd{\begin{definition}}
\def\ed{\end{definition}}
\def\bea{\begin{eqnarray}}
\def\eea{\end{eqnarray}}
\def\bem{\begin{enumerate}}
\def\eem{\end{enumerate}}
\def\bex{\begin{example}}
\def\eex{\end{example}}
\def\bim{\begin{itemize}}
\def\eim{\end{itemize}}
\def\bl{\begin{lemma}}
\def\el{\end{lemma}}
\def\bpf{\begin{proof}}
\def\epf{\end{proof}}
\def\bpp{\begin{proposition}}
\def\epp{\end{proposition}}
\def\bqu{\begin{question}}
\def\equ{\end{question}}
\def\br{\begin{remark}}
\def\er{\end{remark}}
\def\bt{\begin{theorem}}
\def\et{\end{theorem}}

\def\btb{\begin{tabular}}
\def\etb{\end{tabular}}
\newcommand{\nc}{\newcommand}

\def\a{\alpha}

\def\t{\theta}

\def\o{\omega}

\def\G{\Gamma}
\def\D{\Delta}

\def\S{\Sigma}

\def\O{\Omega}

 \nc{\bA}{{\bf A}} \nc{\bB}{{\bf B}} \nc{\bC}{{\bf C}}
 \nc{\bD}{{\bf D}} \nc{\bE}{{\bf E}} \nc{\bF}{{\bf F}}
 \nc{\bG}{{\bf G}} \nc{\bH}{{\bf H}} \nc{\bI}{{\bf I}}
 \nc{\bJ}{{\bf J}} \nc{\bK}{{\bf K}} \nc{\bL}{{\bf L}}
 \nc{\bM}{{\bf M}} \nc{\bN}{{\bf N}} \nc{\bO}{{\bf O}}
 \nc{\bP}{{\bf P}} \nc{\bQ}{{\bf Q}} \nc{\bR}{{\bf R}}
 \nc{\bS}{{\bf S}} \nc{\bT}{{\bf T}} \nc{\bU}{{\bf U}}
 \nc{\bV}{{\bf V}} \nc{\bW}{{\bf W}} \nc{\bX}{{\bf X}}
 \nc{\bZ}{{\bf Z}}

\nc{\cA}{{\cal A}} \nc{\cB}{{\cal B}} \nc{\cC}{{\cal C}}
\nc{\cD}{{\cal D}} \nc{\cE}{{\cal E}} \nc{\cF}{{\cal F}}
\nc{\cG}{{\cal G}} \nc{\cH}{{\cal H}} \nc{\cI}{{\cal I}}
\nc{\cJ}{{\cal J}} \nc{\cK}{{\cal K}} \nc{\cL}{{\cal L}}
\nc{\cM}{{\cal M}} \nc{\cN}{{\cal N}} \nc{\cO}{{\cal O}}
\nc{\cP}{{\cal P}} \nc{\cQ}{{\cal Q}} \nc{\cR}{{\cal R}}
\nc{\cS}{{\cal S}} \nc{\cT}{{\cal T}} \nc{\cU}{{\cal U}}
\nc{\cV}{{\cal V}} \nc{\cW}{{\cal W}} \nc{\cX}{{\cal X}}
\nc{\cZ}{{\cal Z}}

\nc{\hA}{{\hat{A}}} \nc{\hB}{{\hat{B}}} \nc{\hC}{{\hat{C}}}
\nc{\hD}{{\hat{D}}} \nc{\hE}{{\hat{E}}} \nc{\hF}{{\hat{F}}}
\nc{\hG}{{\hat{G}}} \nc{\hH}{{\hat{H}}} \nc{\hI}{{\hat{I}}}
\nc{\hJ}{{\hat{J}}} \nc{\hK}{{\hat{K}}} \nc{\hL}{{\hat{L}}}
\nc{\hM}{{\hat{M}}} \nc{\hN}{{\hat{N}}} \nc{\hO}{{\hat{O}}}
\nc{\hP}{{\hat{P}}} \nc{\hR}{{\hat{R}}} \nc{\hS}{{\hat{S}}}
\nc{\hT}{{\hat{T}}} \nc{\hU}{{\hat{U}}} \nc{\hV}{{\hat{V}}}
\nc{\hW}{{\hat{W}}} \nc{\hX}{{\hat{X}}} \nc{\hZ}{{\hat{Z}}}

\def\dim{\mathop{\rm Dim}}

\def\max{\mathop{\rm max}}

\def\SL{{\mbox{\rm SL}}}
\def\gsl{{\mbox{$\mathfrak {sl}$}}}
\def\SU{{\mbox{\rm SU}}}

\def\Det{{\mbox{\rm Det}}}

\def\ox{\otimes}

\def\sue{\subseteq}

\newcommand{\ket}[1]{|#1\rangle}

\newcommand{\njp}{New. J. Phys.~}

\begin{document}
\title{Proof of the Gour-Wallach conjecture}

\author{Lin Chen}
\affiliation{Department of Pure Mathematics and Institute for
Quantum Computing, University of Waterloo, Waterloo, Ontario, N2L
3G1, Canada} \email{djokovic@uwaterloo.ca}
\email{linchen0529@gmail.com (Corresponding~Author)}

\def\Dbar{\leavevmode\lower.6ex\hbox to 0pt
{\hskip-.23ex\accent"16\hss}D}
\author {{ Dragomir {\v{Z} \Dbar}okovi{\'c}}}
\affiliation{Department of Pure Mathematics and Institute for
Quantum Computing, University of Waterloo, Waterloo, Ontario, N2L
3G1, Canada} \email{djokovic@uwaterloo.ca}

\begin{abstract}
The absolute value of the hyperdeterminant of four qubits is a
useful measure of genuine entanglement. We prove a recent conjecture of Gour and Wallach describing the pure maximally entangled four-qubit states with respect to this measure.
\end{abstract}

\date{ \today }




\maketitle


\section{Introduction}

Maximal entanglement is an important resource in quantum information science. The Bell state is the maximally entangled two-qubit pure state. It contains one entanglement bit (ebit) measured by the von Neumann entropy. Quantum teleportation \cite{bbc93} requires the cost of one ebit and cannot be faithfully carried out by non-maximally entangled states. To make non-maximally entangled states useful for teleportation, remote state preparation \cite{bhl05} and some other quantum-information tasks, they are converted into Bell states by the local operations and classical communications (LOCC). This process is known as the entanglement purification or distillation \cite{bbp96PRL}. For bipartite pure states of higher dimensions, the maximally entangled state \cite{comment} can be used to create any state under LOCC due to  the majorization criterion \cite{nielsen99}. So maximally entangled states are the universal generators for quantum entanglement.

It is thus expected that the multipartite maximally entangled states
will play a similar role in the multipartite systems. The
multipartite entanglement is a more universally operational quantum
resource than the bipartite entanglement. For this purpose we need
to characterize the maximally entangled states in multiqubit
systems. Unlike the bipartite case, the multipartite maximally
entangled states depend on the choice of the multipartite
entanglement measure. It is conjectured that any multiqubit
maximally entangled state has maximally mixed state as the one-party
reduced density operator when the entanglement is measured by the
sum of the negativity \cite{vw02} over all inequivalent bipartitions
\cite{ac13,bss05}. Under a similar condition, the maximal
entanglement with respect to the 4-tangle has been characterized
\cite{gw10}. On the other hand, the W state is the maximally
entangled three-qubit state \cite{cxz10} with respect to the
geometric measure of entanglement, which is a distance-like
entanglement measure \cite{zch10,wg03,amm10}. In this paper, we
explicitly characterize the four-qubit maximally entangled states
(see Eq. \eqref{eq:MaxEnt}) under the absolute value of the
hyperdeterminant introduced in \cite{gkz94,M03}. In the case of
two-qubit and three-qubit pure states, after normalization, the
hyperdeterminant becomes the concurrence and 3-tangle, respectively.
In the four-qubit case, the hyperdeterminant is an invariant
homogeneous polynomial of degree 24, which yields an entanglement
measure for 4-qubit genuine entanglement \cite{gw12}. Our work
mainly confirms a conjecture proposed in \cite{gw12}, to which we
shall refer as the Gour-Wallach conjecture throughout the paper.

Many multiqubit states, such as the ten-photon
Greenberger-Horne-Zeilinger state \cite{gao10}, four-photon W state, and six-photon Dicke states \cite{pct09} have been experimentally realized in recent years. We expect that the maximally entangled 4-qubit state \eqref{eq:MaxEnt} may  be also realized by using the present lab techniques.

In Sec. \ref{sec:pre} we provide some background information on the polynomial invariants and in particular on the hyperdeterminant, and state the Gour-Wallach conjecture ( Conjecture \ref{cj:G-W}). The proof of the conjecture is given in Sec. \ref{sec:proof}. In Sec. \ref{sec:discussion}
we conclude our findings, state an open problem related to
the above conjecture, and make a comment on the generators of the
algebra of symmetric polynomial invariants.

\section{\label{sec:pre} Statement of the conjecture}

We denote by $\SL$ the direct product $\SL_2^{\times4}$ of four
copies of $\SL_2(\bC)$. It is well known (see e.g. \cite{LT03}) that the algebra of polynomial $\SL$-invariants of four qubits is a polynomial algebra, $\cA$, in four variables. The four homogeneous generators of $\cA$ have degrees 2,4,4 and 6. We enlarge $\SL$ by including the group $S_4$ of permutations of four qubits, and obtain the semidirect product 
$\SL^*:=\SL\rtimes S_4$. The algebra of polynomial 
$\SL^*$-invariants, $\cB$, is also a polynomial algebra
in four variables. The four homogeneous generators of $\cB$ have
degrees 2,6,8 and 12, see \cite{CD06,gw12}. The vectors $\psi$ in
the Hilbert space $\cH$ of four qubits can be identified with the
$2\times2\times2\times2$ complex matrices. There is a generalization of determinant to these four-dimensional matrices called {\em hyperdeterminant}, see \cite[Chapter 14]{gkz94} and \cite{M03}, which is a homogeneous $\SL^*$-invariant of degree 24. We shall denote it by $\Det(\psi)$.

All states in this note will be normalized. Let us introduce the subspace $A\sue\cH$ with basis
 \bea
\ket{u_0} &=& \frac{1}{2}\left(
\ket{0000}+\ket{0011}+\ket{1100}+\ket{1111} \right), \\
\ket{u_1} &=& \frac{1}{2}\left(
\ket{0000}-\ket{0011}-\ket{1100}+\ket{1111} \right), \\
\ket{u_2} &=& \frac{1}{2}\left(
\ket{0101}+\ket{0110}+\ket{1001}+\ket{1010} \right), \\
\ket{u_3} &=& \frac{1}{2}\left(
\ket{0101}-\ket{0110}-\ket{1001}+\ket{1010} \right).
 \eea
We write an arbitrary vector $z\in A$ as
 \bea \label{eq:z}
z=\sum_j z_j\ket{u_j},\quad z_j\in\bC.
 \eea
The polynomial $\Det(\psi)$ is usually defined only up to a nonzero constant factor. We use the same normalization as the one adopted in \cite{gw12}. It is specified by the restriction $\Det|_A$, which is given by the formula
 \bea \label{eq:Det}
\Det(z)=\prod_{0\le j<k\le3} (z_j^2-z_k^2)^2,\quad z\in A.
 \eea
(This is not the standard normalization, in which the coefficients of $\Det(\psi)$, considered as a polynomial in the
the $2\times2\times2\times2$ matrix entries, are relatively prime integers.)

We shall denote the unit sphere of $\cH$ by $\Sigma$, and we set $\Sigma_A=\Sigma\cap A$. Gour and Wallach have proposed recently \cite{gw12} to use the absolute value of the hyperdeterminant as a measure of genuine four-qubit entanglement. The maximally entangled states, with respect to this measure, are of special interest and they have proposed the following conjecture concerning these special
states.

 \bcj \label{cj:G-W}
(a) The maximum of $|\Det(\psi)|$ over all states
$\ket{\psi}\in\cH$ is reached at the state
 \bea \label{eq:MaxEnt}
\ket{L}=\frac{1}{\sqrt{3}} \left( \ket{u_0}+\omega\ket{u_1}
+\omega^*\ket{u_2} \right), \quad \omega=e^{i\pi/3}.
 \eea

(b) Up to local unitary (LU) transformations, $\ket{L}$ is the unique state with this property.
 \ecj

By using the formula \eqref{eq:Det} we obtain that
$\Det(L)=-3^{-9}$. By replacing $\o$ by $\o^2$ in the above
expression for $\ket{L}$ we obtain a new state, $\ket{L'}$
\cite{gw10}. It is easy to verify that $\ket{L}$ and $\ket{L'}$ are LU-equivalent. They have been widely studied in
\cite{os05,love07,gw10,gw12}. It is known that, up to local unitary transformations, $\ket{L}$ is the only state that maximizes the average Tsallis $\a$-entropy of entanglement for all $\a>2$ \cite{gw10}.

Let us recall from \cite{gw12} the definition and some properties of the generic set $\O$. By definition, $\O$ is the set of all vectors $\psi\in\cH$ such that $\dim(\SL\cdot\psi)=12$. We have
$\O=\SL\cdot\O_A$ where $\Omega_A:=\Omega\cap A$. Moreover
$\O=\{\psi\in\cH:\Det(\psi)\ne0\}$, and so $\O$ is an open dense
subset of $\cH$.

\section{\label{sec:proof} Proof of the conjecture}

We shall first study the restriction $\Det|_A$. Let
$z=(z_0,z_1,z_2,z_3)$ be a quadruple of complex variables $z_j=r_j
e^{i\theta_j}$, where $r_j\ge0$ and $\theta_j\in\bR$. We denote by
$f$ the basic antisymmetric polynomial in these variables, i.e.,
 \bea \label{eq:f}
f(z)=\prod_{0\le j<k\le3} (z_j-z_k).
 \eea
We shall view $z_j$ as a function of two real variables $r_j$ and $\theta_j$, and so $f$ is a function of eight real variables.
We denote by $U_A$ the open region in $A$ defined by the condition $f(z)\ne0$, i.e., $z_j\ne z_k$ whenever $j\ne k$. Thus if $z\in U_A$ then at most one of the $z_j$ may vanish.

Note that on $A$ we have $\Det(z)=f(z_0^2,\ldots,z_3^2)^2$, i.e., $\Det=(f\circ Q)^2$ where $Q$ is the squaring map
$(z_0,\ldots,z_3)\to(z_0^2,\ldots,z_3^2)$. It follows that
$\O_A=\{z\in A:Q(z)\in U_A\}$. Thus, the maximization problem for $|\Det(z)|$ on $\Sigma_A$ reduces to the problem of maximizing $|f(z)|^2$ subject to the constraint
 \bea \label{eq:constraint}
r_1+r_2+r_3+r_4=1.
 \eea
Let $\D$ denote the closed subset of $A$ defined by this equation.
As $\Det(z)=0$ at the points $z\in\Sigma_A\setminus\O_A$, the maximum of $|\Det|$ on $\Sigma_A$ must be reached at some points $z\in\O_A$. In that case $Q(z)\in U_A$ and $|f|^2$ reaches its maximum on $\D$ at the point $Q(z)$.

Our goal is to show that at all points $z\in\D$ where $|f|^2$
has local maximum on $\D$, we have $|f(z)|^2\le3^{-9}$ and at the same time identify the points at which the equality holds.
Note that $|f|^2$, Eq. \eqref{eq:constraint} and the region $U_A$ are all invariant under permutation of the varables $z_j$ and the multiplication of all $z_j$ by the same phase factor. Therefore the set of local maxima points that we are looking for is also invariant under these transformations and we shall use this fact to simplify the problem.

We shall first treat the case when all $r_j>0$. Note that for
$z\in U_A$ we have
 \bea \label{eq:der-r}
\frac{1}{f} \frac{\partial f}{\partial r_j}=w_j
 \eea
and
 \bea \label{eq:der-t}
\frac{1}{f} \frac{\partial f}{\partial\theta_j}=ir_j w_j
 \eea
where
 \bea \label{eq:w_j}
w_j=e^{i\theta_j} \sum_{k\ne j} \frac{1}{z_j-z_k}.
 \eea
Note also that
 \bea \label{eq:sum-w}
\sum_{j=0}^3 e^{-i\theta_j} w_j=0.
 \eea

Unless stated otherwise, we shall assume that $z\in\D$ is a point
where $|f|^2$ reaches its maximum on $\D$. Then Eqs.
\eqref{eq:der-t} show that the numbers $w_j$ must be real. For $j\ne
k$ we can replace $r_j$ and $r_k$ with $r_j+t$ and $r_k-t$, where
$t$ is an auxiliary real variable. Note that the constraint equation
remains satisfied when $|t|$ is small. Since $z$ is a critical
point, Eqs. \eqref{eq:der-r} imply that
 \bea
\frac{\partial f}{\partial r_j}-\frac{\partial f}{\partial r_k}=(w_j-w_k)f,
 \eea
and we deduce that numbers $w_j-w_k$ must be purely imaginary. On the other hand the numbers $w_j$ are real, and so we  must have
$w_j-w_k=0$. Thus we have shown that $w_0=w_1=w_2=w_3\in\bR$.

Assume that $w_0=0$. Then $w_j=0$ for all $j$, i.e., we have
 \bea \label{eq:rat-functions}
\sum_{k\ne j} \frac{1}{z_j-z_k}=0,\quad j=0,1,2,3.
 \eea
By simplifying the first three of these equations we obtain the
system
 \bea
&& 3z_0^2-2z_0(z_1+z_2+z_3)+(z_1z_2+z_1z_3+z_2z_3)=0, \notag\\
&& 3z_1^2-2z_1(z_0+z_2+z_3)+(z_0z_2+z_0z_3+z_2z_3)=0, \notag\\
&& 3z_2^2-2z_2(z_0+z_1+z_3)+(z_0z_1+z_0z_3+z_1z_3)=0. \notag
 \eea
As the $z_j$ are pairwise distinct, these three equations lead to a contradiction. We conclude that $w_0\ne0$.

Consequently, Eq. \eqref{eq:sum-w} implies that
 \bea \label{eq:sum-theta}
e^{i\theta_0}+e^{i\theta_1}+e^{i\theta_2}+e^{i\theta_3}=0.
 \eea
A simple geometric argument shows that we may assume that
$r_0=\max_j r_j$ and that
 \bea \label{eq:angles}
\t_0=\t,~ \t_1=\pi-\t,~ \t_2=\pi+\t,~ \t_3=-\t
 \eea
for some $\t\in[0,\pi/4]$. By plugging in these expressions into Eq. \eqref{eq:w_j}, we obtain the formulae
 \bea
w_0 &=& \frac{1}{r_0+r_2}+\frac{1}{r_0+r_1u^{-1}}
+\frac{1}{r_0-r_3u^{-1}}, \label{eq:w_0} \\
w_1 &=& \frac{1}{r_1+r_3}+\frac{1}{r_1+r_0 u}
+\frac{1}{r_1-r_2 u}, \label{eq:w_1} \\
w_2 &=& \frac{1}{r_0+r_2}+\frac{1}{r_2-r_1u^{-1}}
+\frac{1}{r_2+r_3u^{-1}}, \label{eq:w_2} \\
w_3 &=& \frac{1}{r_1+r_3}+\frac{1}{r_3-r_0 u}
+\frac{1}{r_3+r_2 u}, \label{eq:w_3}
 \eea
where $u=e^{2i\theta}$. Since the $w_j$ are real, we have
 \bea
\left(r_1|r_0-r_3u|^2-r_3|r_0+r_1u|^2\right)\sin2\t&=&0,\notag\\
\left(r_0|r_1-r_2u|^2-r_2|r_1+r_0u|^2\right)\sin2\t&=&0,\notag\\
\left(r_1|r_2+r_3u|^2-r_3|r_2-r_1u|^2\right)\sin2\t&=&0,\notag\\
\left(r_0|r_3+r_2u|^2-r_2|r_3-r_0u|^2\right)\sin2\t&=&0.\notag
 \eea

If $\t>0$ then the above four equations imply that
 \bea
4r_0r_1r_2r_3 \cos 2\t
&=& r_2(r_1-r_3)(r_0^2-r_1r_3) \notag \\
&=& r_3(r_0-r_2)(r_1^2-r_0r_2) \notag \\
&=& r_0(r_1-r_3)(r_1r_3-r_2^2) \notag \\
&=& r_1(r_0-r_2)(r_0r_2-r_3^2). \label{eq:real-w}
 \eea

Assume first that $\t=0$ and thus $u=1$. From $w_0=w_1$ and
$w_0=w_2$ we obtain that
 \bea
 (r_0-r_1+r_2-r_3)((r_0r_3+r_1r_2)-(r_0r_2+r_1r_3)
 &+&
 \notag\\
 2(r_0r_1+r_2r_3)) &=& 0,  \notag\\
 (r_0+r_1-r_2-r_3)((r_0r_3+r_1r_2)+2(r_0r_2+r_1r_3)
 &-&
 \notag\\
 (r_0r_1+r_2r_3)) &=& 0. \notag
 \eea
Since $z_0\ne z_3$ we have $r_0\ne r_3$ and so either
$r_0-r_1+r_2-r_3=(r_0r_3+r_1r_2)+2(r_0r_2+r_1r_3)
-(r_0r_1+r_2r_3)=0$ or
$r_0+r_1-r_2-r_3=(r_0r_3+r_1r_2)-(r_0r_2+r_1r_3)
+2(r_0r_1+r_2r_3)=0$. Since $r_0+r_1+r_2+r_3=1$, in both cases
we obtain that $|f(z)|^2=2^{-16}<3^{-9}$.

Next assume that $\t=\pi/4$ and so $\cos2\t=0$. If $r_1>r_3$ then
$r_0r_2=r_1^2=r_3^2$ and $r_0>r_1$. A computation shows that
$w_3-w_0=3(r_0-r_1)^2/2r_1(r_0^2+r_1^2)>0$. Thus we have a contradiction. We conclude that $r_0=r_2$. The equality
$(r_1-r_3)(r_1r_3-r_2^2)=0$ implies that $r_1=r_3$. If also
$r_0=r_1$ then all $r_j=1/4$ and we have $f(z)=-1/256$.
Otherwise $r_0>r_1$ and
$w_1-w_0=(r_0-r_1)(r_0^2-4r_0r_1+r_1^2)/2r_0r_1(r_0^2+r_1^2)=0$
implies that $r_0=(3+\sqrt{3})/12$ and $r_2=(3-\sqrt{3})/12$.
Then a computation shows that $|f(z)|^2=6^{-6}<3^{-9}$.

Finally assume that $0<\t<\pi/4$. Then Eqs. \eqref{eq:real-w}
imply that $r_0r_2=r_1r_3$, $r_0\ge r_1>r_3\ge r_2$, and
$\cos2\t=(r_0-r_2)(r_1-r_3)/4r_0r_2$. By using the equation
\eqref{eq:constraint} and $r_0r_2=r_1r_3$, we obtain that
 \bea
r_2=\frac{r_1(1-r_0-r_1)}{r_0+r_1},\quad
r_3=\frac{r_0(1-r_0-r_1)}{r_0+r_1}.
 \eea
A tedious computation now shows that $w_0\ne w_1$ and so we have a contradiction.

Next we consider the case when some $z_j=0$. As $z\in U_A$, at most one of the $z_j$ may vanish. Without any loss of generality we may assume that $z_3=0$, and so $z_0z_1z_2\ne0$. We proceed as in the previous case but we have to make some essential changes.
The equations \eqref{eq:der-r}, \eqref{eq:der-t} and
\eqref{eq:w_j} are now valid only for $j=0,1,2$. As $r_3=0$,
$\t_3$ can be chosen arbitrarily and so $w_3$ is not defined.
Consequently, Eq. \eqref{eq:sum-w} is not meaningful but we have the following substitute
 \bea \label{eq:sum w-z}
\sum_{j=0}^2 \left( w_j - \frac{1}{r_j} \right) e^{-i\t_j} &=& 0.
 \eea
The proof of the assertion $w_0=w_1=w_2\in\bR$ remains valid.
In the proof of the claim that $w_0\ne0$ we used Eq.
\eqref{eq:rat-functions} only for $j=0,1,2$, and so this proof remains valid.

Note that the maximum of $|f(z_0,z_1,z_2,0)|$ when
$z_0,z_1,z_2\in\bR$ and $r_0+r_1+r_2=1$ is equal to the maximum of
$x_0x_1x_2(x_0-x_1)(x_0+x_2)(x_1+x_2)$ where $x_0\ge x_1\ge0$,
$x_2\ge0$ and $x_0+x_1+x_2=1$. By using the method of Lagrange
multipliers, it is easy to verify that the latter maximum is equal
to $2^{-8}$. (The maximum occurs at the point $x_0=1/2$,
$x_1=(2-\sqrt{2})/4$, and $x_2=\sqrt{2}/4$.) Hence, we can dismiss the cases where all $z_j$ are real.

We may assume that $r_0\ge r_1\ge r_2$, $\t_0=0$ and at least one of the phase factors $s_1=e^{i\t_1}$ and $s_2=e^{i\t_2}$ is not real. Eq. \eqref{eq:sum w-z} then implies that neither $s_1$ nor $s_2$ is real.

We claim that $r_0=r_1=r_2=1/3$. Since $w_0=w_1=w_2$ the resultants of $w_0-w_1$ and $w_0-w_2$ with respect to the variable $s_1$ and $s_2$ (separately) must vanish. We obtain the equations
 \bea
&& r_2^2(5r_0-r_1-r_2)(r_0+r_1-r_2)s_2^2
 \notag\\
&&-r_0r_2(5r_0^2-3r_1^2+5r_2^2+2r_0r_1+2r_1r_2-14r_0r_2)s_2
\label{eq:res-1} \notag \\
&& \quad +r_0^2(r_0-r_1-r_2)(r_0+r_1-5r_2)=0, \\
&& r_1^2(5r_0-r_1-r_2)(r_0-r_1+r_2)s_1^2
\notag\\
&&-r_0r_1(5r_0^2+5r_1^2-3r_2^2+2r_0r_2+2r_1r_2-14r_0r_1)s_1
\label{eq:res-2} \notag \\
&& \quad +r_0^2(r_0-r_1-r_2)(r_0+r_2-5r_1)=0,
 \eea
respectively. As $s_1,s_2\notin\bR$ and $|s_1|=|s_2|=1$, the leading and constant terms must be equal in each of these two equations. Thus we obtain the following two equations
 \bea
&& (r_0^3+4r_0r_1r_2+r_2^3-(r_0+r_2)(5r_0r_2+r_1^2))\notag \\
&& \cdot (r_0-r_2)=0, \label{eq:mod-1} \\
&& (r_0^3+4r_0r_1r_2+r_1^3-(r_0+r_1)(5r_0r_1+r_2^2))\notag \\
&& \cdot (r_0-r_1)=0. \label{eq:mod-2}
 \eea
If $r_0=r_1$ then Eq. \eqref{eq:mod-1} implies that $r_0=r_2$ and so our claim holds. Assume now that $r_0>r_1$.
After dropping the factors $r_0-r_2$ and $r_0-r_1$ from the left hand sides of the above equations, the resultant with respect to
$r_0$ of the two remaining polynomials is equal to $288r_1r_2(r_1-r_2)^3(r_1+r_2)^4$. Thus, we obtain that $r_1=r_2$, and so
$r_0=1-2r_1$ and $r_1<1/3$. Eq. \eqref{eq:mod-1} now gives that
$r_1=(9-\sqrt{33})/24$. However, then the roots of Eq.
\eqref{eq:res-1} are real and we have a contradiction. Thus,
our claim is proved.

After setting $r_0=r_1=r_2=1/3$ in Eqs. \eqref{eq:res-1} and
\eqref{eq:res-2}, we deduce that $s_1,s_2\ne1$ are cube roots of 1. As $z_1\ne z_2$ we have $s_1\ne s_2$.

To summarize, we have proved the following lemma.
 \bl \label{le:A-ineq}
The inequality $|f(z)|^2\le3^{-9}$ holds for all $z\in A$ such that $\sum_j |z_j|=1$. The equality holds if and only if exactly one $z_k=0$ while the other three $z_j$ form vertices of an equilateral triangle inscribed in the circle of radius $1/3$
and centered at the origin 0.
 \el

We now shift our focus from $\Det|_A$ to $\Det$ itself. The following lemma plays a crucial role.

 \bl \label{le:NejNorma}
Let $z\in A$ and let $\cO$ denote the $\SL$-orbit through $z$. Then $\|z\| \le \|v\|$ for all points $v\in\cO$ and the equality holds if and only if $v\in\SU\cdot z$, where
$\SU:=\SU(2)^{\times4}$ (a maximal compact subgroup of $\SL$).
 \el
 \bpf
We may assume that $z\ne0$. The function $l_\cO:\cO\to\bR$ whose
value at any $v\in\cO$ is equal to the norm $\|v\|$ is a smooth
function. Let $\gsl=\gsl(2)^{\times4}$ denote the Lie algebra of the
group $\SL$. By using Maple we have verified that the vector $z$ is
orthogonal to the tangent space $\gsl\cdot z$ of $\cO$ at the point
$z$. This means that $z$ is a critical point of the function
$l_\cO$. By the Kempf-Ness theorem (see \cite[Theorem 6.18]{PopVin}
or \cite[Appendix A]{gw10}) the orbit $\cO$ is closed, the function
$l_\cO$ has minimum at the point $z$, and all critical points of
$l_\cO$ correspond to a minimum and constitute a single $\SU$-orbit.
This completes the proof.
 \epf

Now we can prove the conjecture. Let $\psi\in\S$ be an arbitrary state. If $\psi\notin\O$ then $\Det(\psi)=0$, so we may assume that $\psi\in\O$. Any such $\psi$ can be written as
$\psi=g\cdot z$ for some $g\in\SL$ and some $z\in\O_A$. Since $\Det$ is a homogeneous polynomial of degree 24 which is $\SL$-invariant, by using Lemma \ref{le:A-ineq} and Lemma
\ref{le:NejNorma} we have
 \bea \notag
|\Det(\psi)| &=& |\Det(g\cdot z)| \\
&=& |\Det(z)| \notag\\
&=& \|z\|^{24} |\Det(z/\|z\|)| \notag\\
&\le& 3^{-9} \|z\|^{24} \notag\\
&\le& 3^{-9} \|g\cdot z\|^{24} \notag\\
&=& 3^{-9}. \label{eq:Nej-psi}
 \eea
Hence, part (a) of the conjecture is proved.

In order to prove part (b), assume that $|\Det(\psi)|=3^{-9}$. Then the inequality \eqref{eq:Nej-psi} implies that $\|z\|=1$  and by Lemma \ref{le:NejNorma} we have $\psi\in\SU\cdot z$.
(See also \cite[Proposition 17]{gw10}.) Hence, part (b) of the conjecture follows from the following lemma.

 \bl
 \label{le:LU}
All states $\ket{z}=\sum^3_{j=0} z_j\ket{u_j}\in A$ such that
$|\Det(z)|=3^{-9}$ are LU-equivalent to each other.
 \el
 \bpf
First we claim that all permutations of the $\ket{u_i}$,
$i=0,1,2,3$, can be performed by LU-transformations.
Consider the LU-operators:
 \bea
 U_0 &=& \left(
           \begin{array}{cc}
             1 & 0 \\
             0 & -i \\
           \end{array}
         \right)
   \ox
   \left(
           \begin{array}{cc}
             1 & 0 \\
             0 & -i \\
           \end{array}
         \right)
   \ox
   \left(
           \begin{array}{cc}
             1 & 0 \\
             0 & i \\
           \end{array}
         \right)
   \ox
   \left(
           \begin{array}{cc}
             1 & 0 \\
             0 & i \\
           \end{array}
         \right)
  \notag \\
   U_1 &=& \frac{1}{4}
          \left(
           \begin{array}{cr}
             1 & 1 \\
             1 & -1 \\
           \end{array}
         \right)
   \ox
   \left(
           \begin{array}{cr}
             1 & 1 \\
             1 & -1 \\
           \end{array}
         \right)
   \ox
   \left(
           \begin{array}{cr}
             1 & 1 \\
             1 & -1 \\
           \end{array}
         \right)
   \ox
   \left(
           \begin{array}{cr}
             1 & 1 \\
             1 & -1 \\
           \end{array}
         \right)
  \notag \\
   U_2 &=& \left(
           \begin{array}{cc}
             1 & 0 \\
             0 & i \\
           \end{array}
         \right)
   \ox
   \left(
           \begin{array}{cr}
             1 & 0 \\
             0 & -i \\
           \end{array}
         \right)
   \ox
   \left(
           \begin{array}{cr}
             1 & 0 \\
             0 & -i \\
           \end{array}
         \right)
   \ox
   \left(
           \begin{array}{cc}
             1 & 0 \\
             0 & i \\
           \end{array}
         \right).
         \notag
 \eea
It is easy to verify that $U_i$ interchanges $\ket{u_i}$ and $\ket{u_{i+1}}$ and fixes the other two $\ket{u_j}$. Thus our
claim is proved. Hence, by Lemma \ref{le:A-ineq}, we may assume that $z_3=0$ and $(z_0^2,z_1^2,z_2^2)=\frac13(1,\o^2,\o^4)$ where $\o=e^{i\pi/3}$. If $z_0=-1/\sqrt{3}$ we can multiply $\ket{z}$ with the phase factor $-1$. Thus, we can assume that
$z_0=+1/\sqrt{3}$. There are now only four cases to consider. If $z_1z_2=-1/3$ (there are two such cases) then we can multiply $\ket{z}$ with a suitable phase factor and permute the first three $\ket{u_i}$ to obtain the state $\ket{L}$. For instance, if $z_1=\o/\sqrt{3}$ and $z_2=\o^2/\sqrt{3}$ then we would multiply $\ket{z}$ with $\o^{-1}$. Thus, we may assume that $z_1z_2=1/3$, i.e., $\ket{z}=\ket{L}$ or $\ket{L'}$. As mentioned in the Introduction, $\ket{L}$ and $\ket{L'}$ are LU-equivalent. This completes the proof.
 \epf

\section{conclusion and discussion} \label{sec:discussion}

In this paper we have proved the Gour-Wallach conjecture. Thus, the state \eqref{eq:MaxEnt} maximizes the absolute value of the
hyperdeterminant and, up to local unitary transformations, it is 
the unique state with this property. In this sense it is the maximally entangled state of four qubits.

The first step in our proof of this conjecture was to maximize
$|f(z_0,z_1,z_2,z_3)|^2$ subject to the constraint $\sum
|z_j|=1$, where $f(z_0,z_1,z_2,z_3)$ is the Van der Monde
determinant on the complex variables $z_0,z_1,z_2,z_3$. As an
interesting generalization, we would like to propose the following open problem.

Find the maximum, $\mu_n$, of the absolute value of the Van der Monde determinant
 \bea
V_n(z_0,\ldots,z_{n-1})=\prod_{0\le j<k\le n-1} (z_k-z_j)
 \eea
on the complex variables $z_0,\ldots,z_{n-1}$ subject to the
constraint
 \bea
\sum_{j=0}^{n-1} |z_j|=1.
 \eea

The value of $|V_n(z_0,z_1,\ldots,z_{n-1})|$ at the point where $z_{n-1}=0$ and $z_0,\ldots,z_{n-2}$ are vertices of a regular $(n-1)$-gon with center at the origin and radius $1/(n-1)$, e.g.,
 \bea
z_j=\frac{1}{n-1} e^{2\pi ij/(n-1)}, \quad j=0,1,\ldots,n-2,
 \eea
is equal to $\lambda_n:=(n-1)^{-(n-1)^2/2}$. For $n=2,3,4$ we have $\mu_n=\lambda_n$. The proof for $n=2$ is trivial. For $n=4$ see Lemma \ref{le:A-ineq} and its proof. The case $n=3$ is much easier and can be proved by the same method. If $n=2$ the maximum is also attained at the points $(z_0,z_1)$ with $0<|z_0|\le1$ and $z_1=z_0-z_0/|z_0|$.

However, for $n=7$ we have $\mu_7>\lambda_7$ \cite{philip}.

There are two sets of generators of symmetric polynomial invariants of four qubits which have been proposed recently 
\cite{CD06,gw12}. Recall the algebra of symmetric invariants 
$\cB$ mentioned in the introduction. It is generated by four algebraically independent homogeneous polynomials of degrees $2,6,8,12$. We shall express the generators 
$\cF_1,\cF_3,\cF_4,\cF_6$ of $\cB$ constructed in \cite{gw12} as polynomials in the generators $H,\G,\S,\Pi$ constructed in 
\cite{CD06}. By using the restrictions of the generators to the subspace $A$, one can easily verify that
 \begin{eqnarray*}
\cF_1 &=& 2H, \\
\cF_3 &=& 4(3H^3-4\G), \\
\cF_4 &=& \frac{4}{3}\left( 33H^4-104H\G+40\S \right), \\
\cF_6 &=& \frac{4}{3}\left( 513H^6-3012H^3\G+2180H^2\S \right. \\
&& \quad \left. +488\G^2+480\Pi \right).
 \end{eqnarray*}
We point out that in the formulae given in \cite[Table 6]{CD06} the invariants $H,\G,\S,\Pi$ are evaluated at the generic point of $A$, namely $a\ket{u_0}+d\ket{u_1}+b\ket{u_2}+c\ket{u_3}$.
Thus, one should set $a=z_0$, $b=z_2$, $c=z_3$, $d=z_1$ to
get agreement with Eq. \eqref{eq:z}.

For the hyperdeterminant $\Det$, normalized as in \eqref{eq:Det},
we have the equality
 \begin{eqnarray*}
\Det &=& \frac{64}{27}\left( 4H^3\G^3-4H^6\Pi+3H^4\S^2-6H^5\G\S
\right. \\
&+&48H^3\G\Pi-48H^2\S\Pi-96H\G\S^2-96\G^2\Pi \\
&+& \left. 32\S^3-64\Pi^2 +60H^2\G^2\S-36\G^4 \right).
 \end{eqnarray*}

\
\
\

\section*{Acknowledgments}

LC was mainly supported by MITACS and NSERC. DD was supported in
part by an NSERC Discovery Grant.


\begin{thebibliography}{99}

\bibitem{bbc93} C. H. Bennett, G. Brassard, C. Crepeau, R. Jozsa, A. Peres, and W.
Wootters, \prl {\bf 70}, 1895 (1993).

\bibitem{bhl05} C. H. Bennett, P. Hayden, D. W. Leung, P. W. Shor, and A. Winter, IEEE
Trans. Inform. Theory, {\bf51}, 56-74 (2005).

\bibitem{bbp96PRL} C. H. Bennett, G. Brassard, S. Popescu, B.
Schumacher, J. Smolin, and W. K. Wootters, \prl {\bf 76}, 722
(1996).

\bibitem{comment} The $d$-level maximally entangled state has the
form $\frac{1}{\sqrt d}\sum^{d-1}_{i=0}\ket{ii}$. For $d=2$, it
degenerates to the Bell state.

\bibitem{nielsen99} M. A. Nielsen, \prl {\bf 83}, 436 (1999).

\bibitem{vw02} G. Vidal and R. F. Werner, \pra {\bf65},
032314 (2002).

\bibitem{ac13} L. Arnaud and N. J. Cerf, \pra {\bf87}, 012319
(2013).

\bibitem{bss05} I. D. K. Brown, S. Stepney, A. Sudbery, and S. L. Braunstein, J.
Phys. A 38, 1119 (2005).

\bibitem{gw10} G. Gour and N. R. Wallach, J. Math. Phys. {\bf51}, 112201 (2010).

\bibitem{cxz10} Lin Chen, Ai-Min Xu, and Huangjun Zhu, \pra {\bf82}, 032301 (2010).

\bibitem{amm10} M. Aulbach, D. Markham, and M. Murao, \njp {\bf 12}, 073025 (2010).

\bibitem{zch10} Huangun Zhu, Lin Chen, and M. Hayashi, \njp {\bf12}, 083002 (2010); quant-ph/1002.2511.

\bibitem{wg03} T-C.~Wei and P.~M.~Goldbart,
Phys. Rev. A, {\bf68}, 042307 (2003).

\bibitem{gkz94} I. Gelfand, M. Kapranov, and A. Zelevinsky,
Discriminants, Resultants and Multidimensional Determinants,
Birkh\"auser, Boston, 1994.

\bibitem{M03}
A. Miyake, Phys. Rev. A {\bf67}, 012108 (2003).

\bibitem{gw12} G. Gour and N. R. Wallach, quant-ph/1211.5586 (2012).

\bibitem{gao10} Wei-Bao Gao et al, Nature Physics {\bf6}, 331 - 335 (2010).

\bibitem{pct09} R. Prevedel, G. Cronenberg, M. S. Tame, M. Paternostro, P. Walther,
M. S. Kim, A. Zeilinger, Phys. Rev. Lett. {\bf103}, 020503 (2009).

\bibitem{LT03}
J.-G. Luques and J.-Y. Thibon, Phys. Rev. A {\bf67}, 042303 (2003).

\bibitem{CD06}
O.~Chterental and D.~{\v{Z}. \Dbar}okovi{\'c}.
\newblock Normal forms and tensor ranks of pure states of four qubits.
\newblock In G.~D. Ling, editor, {\em Linear Algebra Research Advances},
  chapter~4, pages 133--167. Nova Science Publishers, 2007.

\bibitem{love07} P. J. Love et al., Quantum Inf. Process. {\bf6}, 187 (2007).

\bibitem{os05} A. Osterloh and J. Siewert, Phys. Rev. A. {\bf72}, 012337 (2005);  D.~{\v{Z}. \Dbar}okovi{\'c}. and A. Osterloh, J. Math. Phys. {\bf50}, 033509 (2009);
A. Osterloh and J. Siewert, e-print arXiv:quant-ph/0908:3818.

\bibitem{PopVin} V. L. Popov and E. B. Vinberg,
\textit{Invariant Theory},
in Algebraic Groups IV, Eds. A.N. Parshin and I.R. Shafarevich,
Encycl. Math. Sciences, vol. 55, Springer-Verlag (1994).

\bibitem{philip} Philip Gibbs, e-mail to the authors (August 8, 2013).


\end{thebibliography}
\end{document}